\documentclass[journal,article,submit,moreauthors,10pt,a4paper]{mdpi} 

\firstpage{1} 
\articlenumber{x}
\doinum{10.3390/------}
\pubvolume{xx}
\pubyear{2016}
\copyrightyear{2016}
\externaleditor{Academic Editor: name}
\history{Received: date; Accepted: date; Published: date}

 \theoremstyle{mdpi}
 \newcounter{thm}
 \newcounter{ex}
 \newcounter{re}
 \theoremstyle{mdpidefinition}

 \newtheorem{Definition}[thm]{Definition}

\newcommand{\argmin}{\mathop{\rm arg~min}\limits}

\usepackage{multirow}
\usepackage{bm} 
\usepackage{amsmath,amssymb}

\newcommand{\Com}[1]{\textcolor{black}{#1}}

\preto{\abstractkeywords}{\nolinenumbers}

\Title{Efficient Algorithms for Searching the Minimum Information Partition in Integrated Information Theory}

\Author{Jun Kitazono $^{1,2}$*, Ryota Kanai $^{1}$ and Masafumi Oizumi $^{1,3}$*}
\AuthorNames{Jun Kitazono, Ryota Kanai and Masafumi Oizumi}

\address{%
$^{1}$ \quad Araya, Inc., Toranomon 15 Mori Building 2-8-10 Toranomon Minato-ku Tokyo, 105-0001, JAPAN; \{kitazono, kanair, oizumi\}@araya.org\\
$^{2}$ \quad Graduate School of Engineering, Kobe University, 1-1 Rokkodai-cho Nada-ku Kobe-shi Hyogo, 657-8501, JAPAN\\
$^{3}$ \quad RIKEN Brain Science Institute, 2-1 Hirosawa Wako City Saitama, 351-0198, JAPAN}

\corres{Correspondence: kitazono@araya.org, oizumi@araya.org; Tel.: +81-3-6550-9977}

\abstract{The ability to integrate information in the brain is considered to be an essential property for cognition and consciousness. Integrated Information Theory (IIT) hypothesizes that the amount of integrated information ($\Phi$) in the brain is related to the level of consciousness. IIT proposes that to quantify information integration in a system as a whole, integrated information should be measured across the partition of the system at which information loss caused by partitioning is minimized, called the Minimum Information Partition (MIP). The computational cost for exhaustively searching for the MIP grows exponentially with system size, making it difficult to apply IIT to real neural data. It has been previously shown that if a measure of $\Phi$ satisfies a mathematical property, submodularity, the MIP can be found in a polynomial order by an optimization algorithm. However, although the first version of $\Phi$ is submodular, the later versions are not. In this study, we empirically explore to what extent the algorithm can be applied to the non-submodular measures of $\Phi$ by evaluating the accuracy of the algorithm in simulated data and real neural data. We find that the algorithm identifies the MIP in a nearly perfect manner even for the non-submodular measures. Our results show that the algorithm allows us to measure $\Phi$ in large systems within a practical amount of time.
}

\keyword{integrated information theory; integrated information\Com{;} minimum information partition; submodularity; Queyranne's algorithm; consciousness.}

\begin{document}

\section{Introduction}\label{sec:Intro}
The brain receives various information from the external world. Integrating this information is an essential property for cognition and consciousness \Com{\cite{Tononi1994}}. In fact, phenomenologically, our consciousness is unified. For example, when we see an object, we cannot experience only its shape independently of its color. Or, we cannot experience only the left half of the visual field independently of the right half. Integrated Information Theory of consciousness (IIT) considers that the unification of consciousness should be realized by the ability of the brain to integrate information \cite{Tononi2004,Tononi2008, Oizumi2014}. That is, the brain has internal mechanisms to integrate information about the shape and color of an object or information of the right and left visual field, and therefore our visual experiences are unified. IIT proposes to quantify the degree of information integration by an information theoretic measure ``integrated information'' and hypothesizes that integrated information is related to the level of consciousness. Although the hypothesis is indirectly supported by experiments which showed the breakdown of effective connectivity in the brain during loss of consciousness \cite{Massimini2005,Casali2013}, only a few studies have directly quantified integrated information in real neural data \cite{Lee2009,Chang2012,Boly2015,Haun2017} because of the computational difficulties described below.

Conceptually, integrated information quantifies the degree of interaction between parts or equivalently, the amount of information loss caused by splitting a system into parts \cite{Balduzzi2008,Oizumi2016}. 
\Com{IIT proposes that integrated information should be quantified between the least interdependent parts so that it quantifies information integration in a system as a whole. 
For example, if a system consists of two independent subsystems, the two subsystems are the least interdependent parts. In this case, integrated information is 0 because there is no information loss when the system is partitioned into the two independent subsystems.} 
Such a critical partition of the system is called the Minimum Information Partition (MIP), where information is minimally lost, or equivalently where integrated information is minimized. In general, searching for the MIP requires an exponentially large amount of computational time because the number of partitions exponentially grows with the arithmetic growth of system size $N$. This computational difficulty hinders the application of IIT to experimental data, despite its potential importance in consciousness research and even in broader fields of neuroscience. 

In the present study, we exploit a mathematical concept called submodularity to resolve the combinatorial explosion of finding the MIP. Submodularity is an important concept in set functions which is analogous to convexity in continuous functions. It is known that an exponentially large computational cost for minimizing an objective function is reduced to the polynomial order if the objective function satisfies submodularity. Previously, Hidaka and Oizumi showed that the computational cost for finding the MIP is reduced to $O(N^3)$ \cite{Hidaka2017} by utilizing Queyranne's submodular optimization algorithm \cite{Queyranne1998}. They used mutual information as a measure of integrated information that satisfies submodularity. 
\Com{The measure of integrated information used in the first version of IIT (IIT 1.0) \cite{Tononi2004} is based on mutual information. Thus, if we consider mutual information as a practical approximation of the measure of integrated information in IIT 1.0, Queyranne’s algorithm can be utilized for finding the MIP. However, the practical measures of integrated information in the later versions of IIT \cite{Barrett2010,Oizumi2016,Oizumi2016PLoS,Tegmark2016} are not submodular.}

In this paper, we aim to extend the applicability of submodular optimization to non-submodular measures of integrated information. We specifically consider the three measures of integrated information; mutual information $\Phi_\mathrm{MI}$ \cite{Tononi2004}, stochastic interaction $\Phi_\mathrm{SI}$ \cite{Ay2001,Ay2015,Barrett2010}, and geometric integrated information $\Phi_\mathrm{G}$ \cite{Oizumi2016}. Mutual information is strictly submodular but the others are not. Oizumi et al. previously showed a close relationship among these three measures \cite{Oizumi2016,Amari2017}. From this relationship, we speculate that Queyranne's algorithm might work well for the non-submodular measures. Here, we empirically explore to what extent Queyranne’s algorithm can be applied to the two non-submodular measures of integrated information by evaluating the accuracy of the algorithm in simulated data and real neural data. We find that Queyranne's algorithm identifies the MIP in a nearly perfect manner even for the non-submodular measures. Our results show that Queyranne's algorithm can be utilized even for non-submodular measures of integrated information and makes it possible to practically compute integrated information across the MIP in real neural data, such as multi-unit recordings used in EEG and ECoG, which typically consist of around 100 channels. 
\Com{Although the MIP was originally proposed in IIT for understanding consciousness, it can be utilized to analyze any system irrespective of consciousness such as biological networks, multi-agent systems, and oscillator networks. Therefore, our work would be beneficial not only for consciousness studies but also to other research fields involving complex networks of random variables.}

This paper is organized as follows. We first explain that the three measures of integrated information, $\Phi_\mathrm{MI}$, $\Phi_\mathrm{SI}$, $\Phi_\mathrm{G}$, are closely related from a unified theoretical framework \cite{Oizumi2016,Amari2017} and there is an order relation among the three measures; $\Phi_\mathrm{MI} \geq \Phi_\mathrm{SI} \geq \Phi_\mathrm{G}$. Next, we compare the \Com{partition} found by Queyranne's algorithm with the MIP found by exhaustive search in randomly generated small networks ($N=14$). 
We also evaluate the performance of Queyranne's algorithm in larger networks ($N\sim 20$ and 50 for $\Phi_\mathrm{SI}$ and $\Phi_\mathrm{G}$, respectively). Since the exhaustive search is intractable, we compare Queyranne's algorithm with a different optimization algorithm called the replica exchange Markov Chain Monte Carlo (REMCMC) method \cite{Swendsen1986replica,Geyer1991markov,Hukushima1996exchange,Earl2005parallel}.
Finally, we evaluate the performance of Queyranne's algorithm in ECoG data recorded in monkeys and investigate the applicability of the algorithm in real neural data.

\section{Measures of integrated information}
\begin{figure}[h]
 \begin{center}
  \includegraphics[width=0.8\hsize]{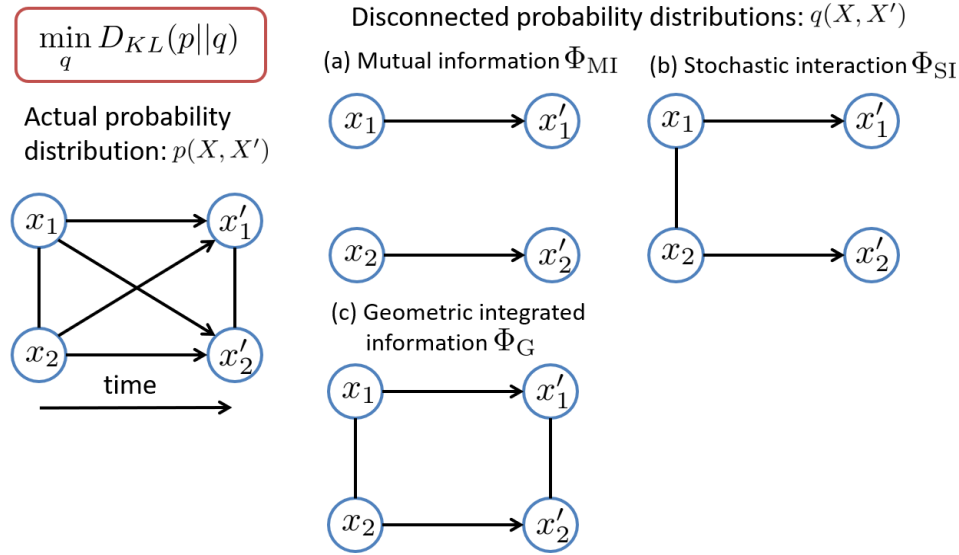} 
  \caption{Measures of integrated information represented by the Kullback-Leibler divergence between the actual distribution $p$ and $q$. (a) Mutual information. (b) Stochastic interaction. (c) Geometric integrated information. \Com{The arrows indicate influences across different time points and the lines without arrowheads indicate influences between elements at the same time.} \Com{This figure is modified from \cite{Oizumi2016}.} \label{fig:phis}}
 \end{center}
\end{figure}

Let us consider a stochastic dynamical system consisting of $N$ elements. We represent the past and present states of the system as $X=(X_1, \ldots , X_N)$ and $X'=(X'_1, \ldots , X'_N)$, respectively.
\Com{In the case of a neural system, the variable $X$ can be signals of multi-unit recordings, EEG, ECoG, and fMRI, etc.}
Conceptually, integrated information is designed to quantify the degree of spatio-temporal interactions between subsystems. The previously proposed measures of integrated information are generally expressed as the Kullback-Leibler divergence between the actual probability distribution $p\left(X,X'\right)$ and a ``disconnected'' probability distribution $q\left(X,X'\right)$ where interactions between subsystems are removed \cite{Oizumi2016}.
\begin{align}
\Phi &= \min_{q} D_{KL}\left(p\left(X,X'\right)||q\left(X,X'\right)\right), \\ 
&= \min_{q} \sum_{x,x'} p\left(x,x'\right) \log \frac{p\left(x,x'\right)}{q\left(x,x'\right)}. \label{eq:phi}
\end{align}
The Kullback-Leibler divergence measures the difference between the probability distributions, and can be interpreted as the information loss when $q\left(X,X'\right)$ is used to approximate $p\left(X,X'\right)$ \cite{Burnham2003}. Thus, integrated information is interpreted as information loss caused by removing interactions. In Eq.~(\ref{eq:phi}), the minimum over $q$ should be taken to find the best approximation of $p$\Com{, while satisfying the constraint that the interactions between subsystems are removed} \cite{Oizumi2016}. 

There are many ways of removing interactions between units, which lead to different disconnected probability distributions $q$, and also different measures of integrated information (Fig.~\ref{fig:phis}). 
\Com{The arrows indicate influences across different time points and the lines without arrowheads indicate influences between elements at the same time.}
Below, we will show that three different measures of integrated information are derived from different probability distributions $q$.

\subsection{Multi (Mutual) information $\Phi_\mathrm{MI}$}

First, consider the following partitioned probability distribution $q$,
\begin{equation}
q\left(X,X'\right)=\prod_i q \left(M_i,M'_i\right),
\end{equation}
where $M_i$ and $M'_i$ are the past and present states of the $i$-th subsystem, respectively. In this model, all of the interactions between the subsystems are removed, i.e., the subsystems are totally independent (Fig.~\ref{fig:phis}~(a)). In this case, the corresponding measure of integrated information is given by 
\begin{equation}
\Phi_\mathrm{MI} = \sum_i H(M_i,M'_i) - H(X,X'), 
\end{equation}
where $H(\cdot,\cdot)$ represents the joint entropy. 
This measure is called total correlation \cite{Watanabe1960} or multi information \cite{Studeny1999}. As a special case when the number of subsystems is two, this measure is simply equivalent to the mutual information between the two subsystems,
\begin{equation}
\Phi_\mathrm{MI} = H(M_1,M'_1) + H(M_2,M'_2) - H(X,X'). \label{eq:phi_MI}
\end{equation}
\Com{The measure of integrated information used in the first version of IIT is based on mutual information but is not identical to mutual information Eq.~(\ref{eq:phi_MI}). The critical difference is that the measures in IIT are based on perturbation and those considered in this study are based on observation. In IIT, a perturbational approach is used for evaluating probability distributions, which attempts to quantify actual causation by perturbing a system into all possible states \cite{Pearl2009,Tononi2004,Balduzzi2008,Oizumi2014}. The perturbational approach requires full knowledge of the physical mechanisms of a system, i.e., how the system behaves in response to all possible perturbations. The measure defined in Eq.~(\ref{eq:phi_MI}) is based on an observational probability distribution that can be estimated from empirical data. Since we aim for the empirical application of our method, we do not consider the perturbational approach in this study.}

\subsection{Stochastic interaction $\Phi_\mathrm{SI}$}

Second, consider the following partitioned probability distribution $q$,
\begin{equation}
q\left(X'|X\right)=\prod_i q\left(M'_i|M_i\right),
\end{equation}
which partitions the transition probability from the past $X$ to the present $X'$ in the whole system into the product of the transition probability in each subsystem. This corresponds to removing the causal influences from $M_i$ to $M'_j$ $(j \neq i)$ as well as the equal time influences \Com{at present} between $M'_i$ and $M'_j$ ($j \neq i$) (Fig.~\ref{fig:phis}~(b)). In this case, the corresponding measure of integrated information is given by
\begin{equation}
\Phi_\mathrm{SI} = \sum_i H(M'_i|M_i) - H(X'|X), 
\end{equation}
where $H(\cdot|\cdot)$ indicates the conditional entropy. This measure was proposed as a practical measure of integrated information by Barrett \& Seth \cite{Barrett2010} following the measure proposed in the second version of IIT (IIT 2.0) \cite{Balduzzi2008}. This measure was also independently derived by Ay as a measure of complexity \cite{Ay2001,Ay2015}. 

\subsection{Geometric integrated information $\Phi_\mathrm{G}$}

Aiming at only the causal influences between parts, Oizumi et al. \cite{Oizumi2016} proposed to measure integrated information with the probability distribution that satisfies
\begin{equation}
q\left(M'_i|X\right)=q\left(M'_i|M_i\right), \forall i \label{eq:cond_ind}
\end{equation}
which means the present state of a subsystem $i$, $M'_i$ only depends on its past state $M_i$.
This corresponds to removing only the causal influences between subsystems while retaining the equal-time interactions between them (Fig.~\ref{fig:phis}~(c)). 
\Com{
The constraint Eq. (\ref{eq:cond_ind}) is equivalent to the Markov condition 
\begin{equation}
q(M’_i, M_i^\mathrm{c}|M_i) = q(M’_i|M_i)q(M_i^\mathrm{c}|M_i), \forall i
\end{equation}
where $M_i^\mathrm{c}$ is the complement of $M_i$. 
This means when $M_i$ is given, $M’_i$ and $M_i^\mathrm{c}$ are independent. 
In other words, the causal interaction between $M_i^\mathrm{c}$ and $M'_i$ is only via $M_i$. 
} 

There is no closed-form expression for this measure in general. However, if the probability distributions are Gaussian, we can analytically solve the minimization over $q$ \Com{(see Appendix \ref{app:Phi_Gauss})}. 

\section{Minimum Information Partition}
In this section, we provide the mathematical definition of Minimum Information Partition (MIP). Then, we formulate the search for MIP as an optimization problem of a set function. 
\Com{The MIP is the partition that divides a system into the least interdependent subsystems so that information loss caused by removing interactions among the subsystems is minimized.} 
The information loss is quantified by the measure of integrated information. 
Thus, the MIP, $\pi_\mathrm{MIP}$, is defined as a partition \Com{\footnote{\Com{Since the minimizer is not necessarily unique, strictly speaking, there could be multiple MIPs.}}} where integrated information is minimized
:
\begin{equation}
  \pi_\mathrm{MIP} := \argmin_{\pi\in \mathcal{P}}\Phi(\pi), \label{eq:def_MIP}
\end{equation}
where $\mathcal{P}$ is a set of partitions. 
In general, $\mathcal{P}$ is the universal set of partitions, including bi-partitions, tri-partitions, and so on. In this study, however, we focus only on bi-partitions for simplicity and computational time. By a bi-partition, a whole system $\Omega$ is divided into a subset $S$ $(S\subset\Omega, S\neq \emptyset)$ and its complement $\bar{S}=\Omega \setminus S$. Since a bi-partition is uniquely determined by specifying a subset $S$, integrated information can be considered as a function of a set $S$, $\Phi(S)$. Finding the MIP is equivalent to finding the subset, $S_\mathrm{MIP}$, that achieves the minimum of integrated information:
\begin{equation}
S_\mathrm{MIP}:=\argmin_{S\subset\Omega, S\neq \emptyset}\Phi(S).
\end{equation}
In this way, the search of the MIP is formulated as an optimization problem of a set function. 

Since the number of bi-partitions for the system with $N$-elements is $2^{N-1}-1$, exhaustive search of the MIP in a large system is intractable. However, by formulating the MIP search as an optimization of a set function as above, we can take advantage of a discrete optimization technique and can reduce computational costs to a polynomial order, as described in the next section.

\section{Submodular optimization}
The submodularity is an important concept in set functions, which is an analogue of convexity in continuous functions \cite{Iwata2008}. When objective functions are submodular, efficient algorithms are available for solving optimization problems. In particular, for symmetric submodular functions, there is a well-known algorithm by Queyranne which minimizes them \cite{Queyranne1998}. We utilize this method for finding the MIP in this study.

\subsection{Submodularity}
Mathematically, the submodularity is defined as follows.
\begin{Definition}[Submodularrity]
Let $\Omega$ be a finite set and $2^\Omega$ its power set. A set function $f: 2^\Omega\rightarrow\mathbb{R}$ is submodular if it satisfies the following inequality for any $S, T \subseteq \Omega$: 
\begin{equation*}
f(S)+f(T) \geq f(S\cup T)+f(S\cap T).
\end{equation*}
Equivalently, a set function $f: 2^\Omega\rightarrow\mathbb{R}$ is submodular if it satisfies the following inequality for any $S, T \subseteq \Omega$ with $S \subseteq T$ and for any $u\in\Omega\setminus T$: 
\begin{equation*}
f(S\cup\{u\})-f(S) \geq f(T\cup\{u\})-f(T).
\end{equation*}
\end{Definition}
The second inequality means that the function increases more when an element is added to a smaller subset than when the element is added to a bigger subset.

\subsection{Queyranne's algorithm}
A set function $f$ is called symmetric if $f(S)=f(\Omega\setminus S)$ for any $S\subseteq \Omega$. Integrated information $\Phi(S)$ computed by bi-partition is a symmetric function, because $S$ and $\Omega\setminus S$ specifies the same bi-partition. If a function is symmetric and submodular, we can find the minimum of the function by Queyranne’s algorithm with $O(N^3)$ function calls \cite{Queyranne1998}.
\subsection{Submodularity in measures of integrated information} \label{seq:sub_phi}
In a previous study, Queyranne's algorithm was utilized to find the MIP when $\Phi_\mathrm{MI}$ is used as the measure of integrated information \cite{Hidaka2017}. As shown previously, $\Phi_\mathrm{MI}$ is submodular \cite{Hidaka2017}. However, the other measures of integrated information are not submodular. In this study, we apply Queyranne's algorithm to non-submodular functions, $\Phi_\mathrm{SI}$ and $\Phi_\mathrm{G}$. When the objective functions are not submodular, Queyranne's algorithm does not necessarily find the MIP. We evaluate how accurately Queyranne's algorithm can find the MIP when it is used for non-submodular measures of integrated information.
There is an order relation among the three measures of integrated information \cite{Oizumi2016},
\begin{equation}
\Phi_\mathrm{G} \leq \Phi_\mathrm{SI} \leq \Phi_\mathrm{MI}.  \label{Eq:order}
\end{equation}
This inequality can be graphically understood from Fig.~\ref{fig:phis}. The more the connections are removed, the larger the corresponding integrated information (the information loss) is.
\Com{That is, $\Phi_\mathrm{G}$ measures only the causal influences between subsystems, $\Phi_\mathrm{SI}$ measures the equal-time interactions between the present states as well as the causal influences between subsystems, and $\Phi_\mathrm{MI}$ measures all the interactions between the subsystems. Thus, $\Phi_\mathrm{SI}$ is closer to $\Phi_\mathrm{MI}$ than $\Phi_\mathrm{G}$ is. This relationship implies that $\Phi_\mathrm{SI}$ would behave more similarly to a submodular measure $\Phi_\mathrm{MI}$ than $\Phi_\mathrm{G}$ does. Thus, one may surmise that Queyranne's algorithm would work  more accurately for $\Phi_\mathrm{SI}$ than for $\Phi_\mathrm{G}$. As we will show in Subsection \ref{Subsec:Accuracy}, this is indeed the case. However, the difference is rather small because Queyranne's algorithm works almost perfectly for both measures, $\Phi_\mathrm{SI}$ and $\Phi_\mathrm{G}$.}

\section{Replica Exchange Markov Chain Monte Carlo Method}
To evaluate the accuracy of Queyranne's algorithm, we compare the \Com{partition} found by Queyranne's algorithm with the MIP found by the exhaustive search when the number of elements $n$ is small enough ($n \lesssim 20$). However, when $n$ is large, we cannot know the MIP because the exhaustive search is unfeasible. To evaluate the performance of Queyranne's algorithm in a large system, we compare it with a different method, the Replica Exchange Markov Chain Monte Carlo (REMCMC) method \cite{Swendsen1986replica,Geyer1991markov,Hukushima1996exchange,Earl2005parallel}. REMCMC, also known as parallel tempering, is a method to draw samples from probability distributions. REMCMC is an improved version of the MCMC methods. Here, we briefly explain how the MIP search problem is represented as a problem of drawing samples from a probability distribution. Details of the REMCMC method are given in Appendix \ref{app:REMCMC}. 

Let us define a probability distribution $p(S;\beta)$ using integrated information $\Phi(S)$ as follows:
\begin{equation}
  p(S;\beta) \propto \exp(- \beta \Phi(S) ), \label{eq:Boltzmann}
\end{equation}
where $\beta (> 0)$ is a parameter called inverse temperature. This probability is higher/lower when $\Phi(S)$ is smaller/larger. The MIP gives the highest probability by definition. If we can draw samples from this distribution, we can selectively scan subsets with low integrated information and efficiently find the MIP, compared to randomly exploring partitions independent of the value of integrated information. \Com{Simple MCMC methods like the Metropolis method, which draw samples from Eq. (\ref{eq:Boltzmann}) with a single value of $\beta$, often suffer from the problem of slow convergence. That is, a sample sequence is trapped in a local minimum and the sample distribution takes time to converge to the target distribution. REMCMC aims at overcoming this problem by drawing samples in parallel from distributions with multiple values of $\beta$ and by continually exchanging the sampled sequences between neighboring $\beta$ (See Appendix \ref{app:REMCMC} for more details).}


\section{Results}
We first evaluated the performance of Queyranne's algorithm in simulated networks. 
\Com{Throughout the simulations below, we consider the case where the variable $X$ obeys a Gaussian distribution for the ease of computation. As shown in Appendix \ref{app:Phi_Gauss}, the measures of integrated information, $\Phi_\mathrm{SI}$ and $\Phi_\mathrm{G}$ can be analytically computed. Note that although $\Phi_\mathrm{SI}$ and $\Phi_\mathrm{G}$ can be computed in principle even when the distribution of $X$ is not Gaussian, it is practically very hard to compute them in large systems because the computation of $\Phi$ involves summation over all possible $X$. Specifically, we consider the first order autoregressive (AR) model,}
\begin{equation}
X' = A X + E,
\end{equation}
where $X$ and $X'$ are present states and past states of a system, $A$ is the connectivity matrix, and $E$ is Gaussian noise. The stationary distribution of this AR model is considered. The stationary distribution of $p(X,X')$ is a Gaussian distribution. The covariance matrix of $p(X,X')$ consists of covariance of $X$, $\Sigma(X)$, and cross-covariance of $X$ and $X'$, $\Sigma(X,X')$. $\Sigma(X)$ is computed by solving the following equation, 
\begin{equation}
\Sigma(X) = A \Sigma(X) A^T + \Sigma(E).
\end{equation}
$\Sigma(X,X')$ is given by
\begin{equation}
\Sigma(X,X') = \Sigma(X) A^T.
\end{equation}
By using these covariance matrices, $\Phi_\mathrm{SI}$ and $\Phi_\mathrm{G}$ are analytically calculated \cite{Oizumi2016} (see Appendix \ref{app:Phi_Gauss}). The details of the parameter settings are described in each subsection.

\subsection{Speed of Queyranne's algorithm compared with exhaustive search}
 

\begin{figure}[h]
 \begin{center}
 \includegraphics[width=1\hsize]{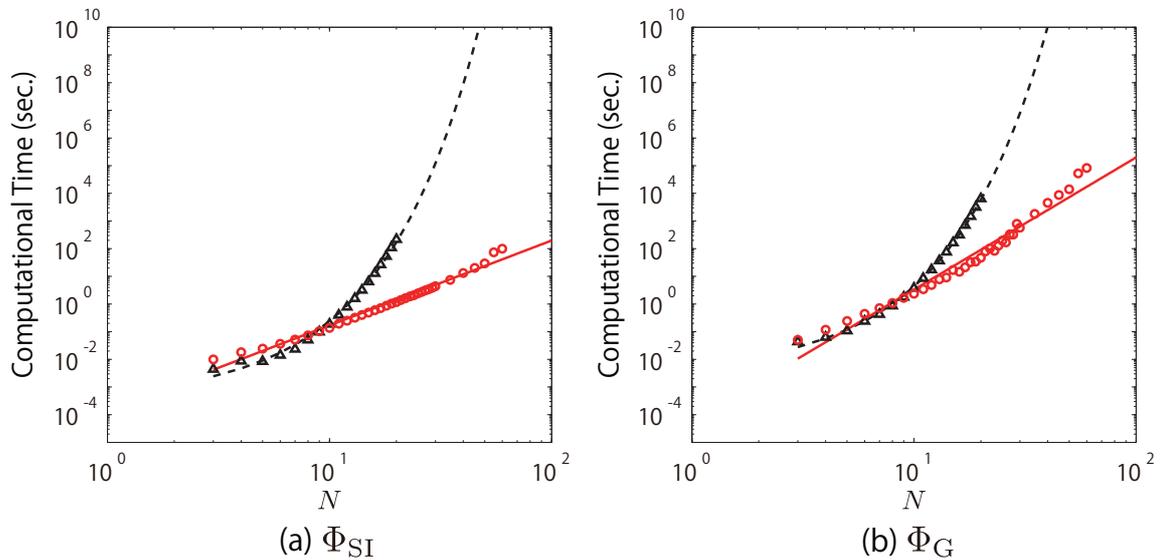}
 \caption{Computational time of Queyranne's algorithm and the Exhaustive search. \Com{The red circles and the red solid lines indicate the computational time of Queyranne's algorithm and their approximate curves ((a) $\log_{10}T=3.066\log_{10}N-3.838$, (b) $\log_{10}T=4.776\log_{10}N-4.255$.) The black triangles and the black dashed lines indicate the computational time of the exhaustive search and their approximate curves ((a) $\log_{10}T=0.2853N-3.468$, (b) $\log_{10}T=0.3132N-2.496$.)}}\label{Fig:Speed}
 \end{center}
\end{figure}

We first evaluated the computation time of Queyranne's algorithm and compared it with that of the exhaustive search when the number of elements $N$ changed. The connectivity matrices $A$ were randomly generated. Each element of the connection matrix $A$ was sampled from a normal distribution with mean 0 and variance $0.01/N$. The covariance of Gaussian noise $E$ was generated from a Wishart distribution $\mathcal{W}(\sigma I, 2N)$ with covariance $\sigma I$ and degrees of freedom $2N$, where $\sigma$ corresponded to the amount of noise $E$ and $I$ was the identity matrix. 
\Com{The Wishart distribution is a standard distribution for symmetric positive-semidefinite matrices \cite{Wishart1928,Bishop2006}. Typically, the distribution is used to generate covariance matrices and inverse covariance (precision) matrices. For more practical details, see for example, \cite{Bishop2006}.} We set $\sigma$ to 0.1. 
\Com{The number of elements $N$ was changed from 3 to 60.}
\Com{All computation times were measured on a machine with an Intel Xeon CPU E5-2680 at 2.70GHz. All the calculations were implemented in MATLAB.}

Figure~\ref{Fig:Speed} (a) shows the results for $\Phi_\mathrm{SI}$. The red circles, which indicate the computational time of Queyranne's algorithm, are fit by the red solid line, $\log_{10}T=3.066\log_{10}N-3.838$. In contrast, the black triangles, which indicate those of the exhaustive search, are fit by the black dashed line, $\log_{10}T=0.2853N-3.468$. This means that the computational time of Queyranne's algorithm increases in polynomial order ($T\propto N^{3.066}$), while that of the exhaustive search exponentially increases ($T\propto1.929^N$). For example, when $N=100$, Queyanne's algorithm takes $\sim 197$ sec while the exhaustive search takes $1.16 \times 10^{25}$ sec. This is in practice impossible to compute even with a \Com{supercomputer}. Similarly, as shown in Fig.~\ref{Fig:Speed} (b), when $\Phi_\mathrm{G}$ is used, \Com{Queyranne's algorithm roughly takes $T\propto N^{4.776}$} while the exhaustive search takes $T\propto 2.057^N$. Note that the reason why the order of the computational time of Queyrannes algorithm for $\Phi_\mathrm{G}$ is higher than that for $\Phi_\mathrm{SI}$ is because the multi-dimensional optimization is needed to compute $\Phi_\mathrm{G}$ (see Appendix \ref{app:Phi_Gauss}).

\subsection{Accuracy of Queyranne's algorithm}\label{Subsec:Accuracy}
\begin{table}[h]
 \begin{center}
 \caption{Accuracy of Queyranne's algorithm}\label{Tab:Accuracy}
  \begin{tabular}{c c  r r r r  r r r r} \toprule
    \multicolumn{2}{c}{Model} & \multicolumn{4}{c}{$\boldsymbol{\Phi_\mathrm{SI}}$} & \multicolumn{4}{c}{$\boldsymbol{\Phi_\mathrm{G}}$} \\ \cmidrule(lr){3-6} \cmidrule(lr){7-10}
    $A$    & $\sigma$ & CR & RA & ER & CORR & CR & RA & ER & CORR \\ \midrule
    \multirow{2}{*}{Normal} & 0.01 & 100\% & 1 & 0 & 1 & 100\% & 1    & 0       & 1 \\ 
                            &  0.1 & 100\% & 1 & 0 & 1 & 100\% & 1    & 0       & 1 \\ \midrule
    \multirow{2}{*}{Block}  & 0.01 & 100\% & 1 & 0 & 1 &  97\% & 1.05 & 2.38e-3 & 0.978 \\ 
                            &  0.1 & 100\% & 1 & 0 & 1 &  97\% & 1.03 & 9.11e-4 & 0.978 \\ \bottomrule
  \end{tabular}
  \end{center}
\end{table}

We evaluated the accuracy of Queyranne's algorithm by comparing the \Com{partition} found by Queyranne's algorithm with the MIP found by exhaustive search. We used $\Phi_\mathrm{SI}$ and $\Phi_\mathrm{G}$ as the measures of integrated information. We considered two different architectures in connectivity matrix $A$ of AR models. The first one was just a random matrix: Each element of $A$ was randomly sampled from a normal distribution with mean 0 and variance $0.01/N$. The other one was a block matrix consisting of $N/2$ by $N/2$ sub-matrices, $A_{ij} (i,j=1, 2)$. Each element of diagonal sub-matrices $A_{11}$ and $A_{22}$ was drawn from a normal distribution with mean 0 and variance $0.02/N$. Off-diagonal sub-matrices $A_{12}$ and $A_{21}$ were zero matrices. The covariance of Gaussian noise $E$ in the AR model was generated from a Wishart distribution $\mathcal{W}(\sigma I, 2N)$. The parameter $\sigma$ was set to 0.1 or 0.01. \Com{The number of elements $N$ was set to 14.}
We randomly generated 100 connectivity matrices $A$ and $\Sigma(E)$ for each setting and evaluated performance using the following four measures. The following measures are averaged over 100 trials. 
\begin{description}
\item[Correct rate (CR)]
Correct rate (CR) is the rate of correctly finding the MIP. 

\item[Rank (RA)]
Rank (RA) is the rank of the partition found by Queyranne's algorithm among all possible partitions. The rank is based on the $\Phi$ values computed at each partition. The partition that gives the lowest $\Phi$ is rank 1. The highest rank is equal to the number of possible bi-partitions, $2^{N-1}$. 

\item[Error ratio (ER)]
Error ratio (ER) is the deviation of the value of integrated information computed across the partition found by Queyranne's algorithm from that computed across the MIP, which is normalized by the mean error computed at all possible partitions. Error ratio is defined by 
\begin{equation}
  \textrm{Error Ratio} = \frac{\Phi_\mathrm{Q}-\Phi_\mathrm{MIP}}{\bar{\Phi}-\Phi_\mathrm{MIP}},
\end{equation}
where $\Phi_\mathrm{MIP}$, $\Phi_\mathrm{Q}$, and $\bar{\Phi}$ are the amount of integrated information computed across the MIP, that computed across the partition found by Queyranne's algorithm, and the mean of the amounts of integrated information computed across all possible partitions, respectively. 

\item[Correlation (CORR)]
Correlation (CORR) is the correlation between the partition found by Queyranne's algorithm and the MIP found by the exhaustive search. Let us represent a bi-partition of $N$-elements as an $N$-dimensional vector $\bm{\sigma}=(\sigma_1,\ldots,\sigma_N)\in\{-1, 1\}^N$, where $\pm 1$ indicates one of the two subgroups.
\Com{The absolute value of the correlation between the vector given by the MIP ($\bm{\sigma}^\mathrm{MIP}$) and that given by the \Com{partition} found by Queyranne's algorithm ($\bm{\sigma}^\mathrm{Q}$) is computed: 
\begin{equation}
  |\textrm{corr}(\bm{\sigma}^\mathrm{MIP}, \bm{\sigma}^\mathrm{Q})| = 
  \left|\frac{\sum_{i=1}^N (\sigma_i^\mathrm{MIP} - \bar{\sigma}^\mathrm{MIP})(\sigma_i^\mathrm{Q} - \bar{\sigma}^\mathrm{Q})}
       {\sqrt{\sum_{i=1}^N (\sigma_i^\mathrm{MIP} - \bar{\sigma}^\mathrm{MIP})^2 \sum_{i=1}^N (\sigma_i^\mathrm{Q} - \bar{\sigma}^\mathrm{Q})^2}}\right|,
\end{equation} 
where $\bar{\sigma}^\mathrm{MIP}$ and $\bar{\sigma}^\mathrm{Q}$ are the means of $\sigma_i^\mathrm{MIP}$ and $\sigma_i^\mathrm{Q}$, respectively.}

\end{description}

The results are summarized in Table~\ref{Tab:Accuracy}. This table shows that, when $\Phi_\mathrm{SI}$ was used, Queyranne's algorithm perfectly found the MIPs for all 100 trials, even though $\Phi_\mathrm{SI}$ is not strictly submodular. Similarly, when $\Phi_\mathrm{G}$ was used, Queyranne's algorithm almost perfectly found the MIPs. The correct rate was 100\% for the normal models and 97\% for the block structured models. Additionally, even when the algorithm missed the MIP, the rank of the partition found by the algorithm was 2 or 3. The averaged rank over 100 trials were 1.03 and 1.05 for the block structured models. Also, the error ratio in error trials were around 0.1 and the average error ratios were very small.
\Com{See Appendix \ref{app:boxplots} for box plots of the values of the integrated information at all the partitions.}
Thus, such miss trials would not affect evaluation of the amount of integrated information in practice. 
\Com{However, in terms of partitions, the partitions found by Queyranne's algorithm in error trials were markedly different from the MIPs. In the block structured model, the MIP for $\Phi_\mathrm{G}$ was the partition that split the system in halves. In contrast, the partitions found by Queyranne's algorithm were one-vs-all partitions.}

In summary, Queyranne's algorithm perfectly worked for $\Phi_\mathrm{SI}$. With regards to $\Phi_\mathrm{G}$, although Queyranne's algorithm almost perfectly evaluated the amount of integrated information, we may need to treat partitions found by the algorithm carefully. This slight difference in performance between $\Phi_\mathrm{SI}$ and $\Phi_\mathrm{G}$ can be explained by the order relation Eq.~(\ref{Eq:order}). $\Phi_\mathrm{SI}$ is closer to the strictly submodular function $\Phi_\mathrm{MI}$ than $\Phi_\mathrm{G}$ is, which we consider to be why Queyranne's algorithm worked better for $\Phi_\mathrm{SI}$ than $\Phi_\mathrm{G}$.

\subsection{Comparison between Queyranne's algorithm and REMCMC}\label{Subsec:QvsREMCMC}
\begin{table}[h]
 \begin{center}
 \caption{Comparison of Queyranne's algorithm with REMCMC ($\Phi_\mathrm{SI}$, $N=50$) }\label{Tab:QvsREMCMC_SI}
  \begin{tabular}{c c  c c c  c c c} \toprule
    \multicolumn{2}{c}{Model} & \multicolumn{3}{c}{\textbf{Winning percentage}} & \multicolumn{3}{c}{\textbf{Number of evaluations of $\Phi$}} \\ \cmidrule(lr){3-5} \cmidrule(lr){6-8}
         &          &             &      &        &             & \multicolumn{2}{c}{REMCMC (mean$\pm$std)} \\ \cmidrule(lr){7-8}
    $A$  & $\sigma$ & Queyranne's & Even & REMCMC & Queyranne's & Converged &Solution found \\ \midrule
    \multirow{2}{*}{Normal}& 0.01 & 0\% & 100\% & 0\% & 41,699 & 274,257$\pm$107,969 &8,172.6$\pm$6,291.0\\ 
                           &  0.1 & 0\% & 100\% & 0\% & 41,699 & 315,050$\pm$112,205 &9,084.9$\pm$7,676.4\\ \midrule
    \multirow{2}{*}{Block} & 0.01 & 0\% & 100\% & 0\% & 41,699 & 308,976$\pm$110,905 &7,305.6$\pm$6,197.0\\
                           &  0.1 & 0\% & 100\% & 0\% & 41,699 & 339,869$\pm$154,161 &4,533.4$\pm$3,004.8\\ \bottomrule
  \end{tabular}
  \end{center}
\end{table}

\begin{table}[h]
 \begin{center}
 \caption{Comparison of Queyranne's algorithm with REMCMC ($\Phi_\mathrm{G}$, $N=20$) }\label{Tab:QvsREMCMC_G}
    \begin{tabular}{c c  c c c  c c c} \toprule
    \multicolumn{2}{c}{Model} & \multicolumn{3}{c}{\textbf{Winning percentage}} & \multicolumn{3}{c}{\textbf{Number of evaluations of $\Phi$}} \\ \cmidrule(lr){3-5} \cmidrule(lr){6-8}
         &          &             &      &        &             & \multicolumn{2}{c}{REMCMC (mean$\pm$std)} \\ \cmidrule(lr){7-8}
    $A$  & $\sigma$ & Queyranne's & Even & REMCMC & Queyranne's & Converged & Solution found \\ \midrule
    \multirow{2}{*}{Normal}& 0.01 & 0\% & 100\% & 0\% & 2,679 & 136,271$\pm$46,624 & 862.4$\pm$776.3\\ 
                           &  0.1 & 0\% & 100\% & 0\% & 2,679 & 122,202$\pm$46,795 & 894.3$\pm$780.2\\ \midrule
    \multirow{2}{*}{Block} & 0.01 & 0\% & 100\% & 0\% & 2,679 & 129,770$\pm$88,483 & 245.2$\pm$194.3\\
                           &  0.1 & 0\% & 100\% & 0\% & 2,679 & 146,034$\pm$61,880 & 443.2$\pm$642.1\\ \bottomrule
  \end{tabular}
  \end{center}
\end{table}
We evaluated the performance of Queyranne's algorithm in large systems where an exhaustive search is impossible. We compared it with the Replica Exchange Markov Chain Monte Carlo Method (REMCMC). We applied the two algorithms to AR models generated similarly as in the previous section. The number of elements was 50 for $\Phi_\mathrm{SI}$ and 20 for $\Phi_\mathrm{G}$, respectively. 
The reason for the difference in $N$ is because $\Phi_\mathrm{G}$ requires much heavier computation than $\Phi_\mathrm{SI}$ (see Appendix \ref{app:Phi_Gauss}). 
We randomly generated 20 connectivity matrices $A$ and $\Sigma(E)$ for each setting. We compared the two algorithms in terms of the amount of integrated information and the number of evaluations of $\Phi$. REMCMC was run until a convergence criterion was satisfied. See Appendix~\ref{app:convergence} for details of the convergence criterion. 

The results are shown in Tables~\ref{Tab:QvsREMCMC_SI} and \ref{Tab:QvsREMCMC_G}. ``Winning percentage'' indicates the fraction of trials each algorithm won in terms of the amount of integrated information at the partition found by each algorithm. 
We can see that the partitions found by the two algorithms exactly matched for all the trials. We consider that the algorithms probably found the MIPs for the following three reasons. First, it is well known that REMCMC can find a minima if it is run for a sufficiently long time in many applications \cite{Earl2005parallel,Pinn1998number,Hukushima2002extended,Nagata2015exhaustive}. Second, the two algorithms are so different that it is unlikely that they both incorrectly identified the same partitions as the MIPs. Third, Queyranne's algorithm successfully finds the MIPs in smaller systems as shown in the previous section. This fact suggests that Queyranne's algorithm worked well also for the larger systems. 
\Com{Note that in the case of $\Phi_\mathrm{G}$, the half-and-half partition is the MIP in the block structured model because $\Phi_\mathrm{G}=0$ under the half-and-half partition. We confirmed that the partitions found by Queyanne’s algorithm and REMCMC were both the half-and-half partition for all the 20 trials. Thus, in the block structured case, it is certain that the true MIPs were successfully found by both algorithms.}

We also evaluated the number of evaluations of $\Phi$ in both algorithms before the end of the computational processes. In our simulations, the computational process of Queyranne's algorithm ended much faster than the convergence of REMCMC. Queyranne's algorithm ends at a fixed number of evaluations of $\Phi$ depending only on $N$. In contrast, the number of the evaluations before the convergence of REMCMC depends on many factors such as the network models, the initial conditions, and pseudo random number sequences. Thus, the time of convergence varies among different trials. Note that by ``retrospectively'' examining the sequence of the Monte Carlo search, the \Com{solutions} turned out to be found at earlier points of the Monte Carlo searches than Queyranne's algorithm (which are indicated as ``\Com{solution} found'' in Tables~\ref{Tab:QvsREMCMC_SI} and \ref{Tab:QvsREMCMC_G}). However, it is impossible to stop the REMCMC algorithm at these points where the solutions were found because there is no way to tell whether these points reach the \Com{solution} until the algorithm is run for enough amount of time.

\subsection{Evaluation with real neural data}\label{Subsec:real}
Finally, to ensure the applicability of Queyranne's algorithm to real neural data, we similarly evaluated the performance with electrocorticogram (ECoG) data recorded in a macaque monkey. The dataset is available at an open database, Neurotycho.org (http://neurotycho.org/). One hundred and twenty-eight channel ECoG electrodes were implanted in the left hemisphere. The electrodes were placed at 5-mm intervals, covering the frontal, parietal, temporal, and occipital lobes, and medial frontal and parietal walls. Signals were sampled at a rate of 1 kHz and down-sampled to 100 Hz for the analysis. The monkey ``Chibi'' was awake with the eyes covered by an eye-mask to restrain visual responses. To remove line noise and artifacts, we performed bipolar re-referencing between nearest neighbor electrode pairs. The number of re-referenced electrodes was 64 in total. 

In the first simulation, we evaluated the accuracy. We extracted a 1-minute length of the signals of the 64 electrodes. 
\Com{Each 1-minute sequence consists of 100 Hz $\times$ 60 sec. $=$ 6000 samples.}
\Com{Then, we randomly selected
14 electrodes 100 times.}
We approximated the probability distribution of the signals with multivariate Gaussian distributions. The covariance matrices were computed with a time window of 1 min and a time step of 10 ms. We applied the algorithms to the 100 randomly selected sets of electrodes and measured the accuracy similarly as in Subsection~\ref{Subsec:Accuracy}. The results are summarized in Table~\ref{Tab:Accuracy_NeuroTycho}. We can see that Queyranne's algorithm worked perfectly for both $\Phi_\mathrm{SI}$ and $\Phi_\mathrm{G}$. 

Next, we compared Queyranne's algorithm with REMCMC. We applied the two algorithms to the 64 re-referenced signals, and evaluated the performance in terms of the amount of integrated information and the number of evaluations of $\Phi$, as in Subsection~\ref{Subsec:QvsREMCMC}. We segmented 15 non-overlapping sequences of 1 minute each, and computed covariance matrices with a time step of 10 ms. We measured the average performance over the 15 sets. Here, we only used $\Phi_\mathrm{SI}$, because $\Phi_\mathrm{G}$ requires heavy computations for 64 dimensional systems. The results are shown in Table~\ref{Tab:QvsREMCMC_SI_NeuroTycho}. We can see that the partitions selected by the two algorithms matched for all 15 sequences. 
In terms of the amount of computation, Queyranne's algorithm ended much faster than the convergence of REMCMC. 

\begin{table}[h]
 \begin{center}
 \caption{Accuracy of Queyranne's algorithm in ECoG data. \Com{Randomly-selected 14 electrodes were used.}}\label{Tab:Accuracy_NeuroTycho}
  \begin{tabular}{r r r r  r r r r} \toprule
    \multicolumn{4}{c}{$\boldsymbol{\Phi_\mathrm{SI}}$} & \multicolumn{4}{c}{$\boldsymbol{\Phi_\mathrm{G}}$} \\ \cmidrule(lr){1-4} \cmidrule(lr){5-8}
    CR & RA & ER & CORR & CR & RA & ER & CORR \\ \midrule
   100\% & 1 & 0 & 1 & 100\% & 1    & 0       & 1 \\ \bottomrule
  \end{tabular}
  \end{center}
\end{table}
\begin{table}[h]
 \begin{center}
 \caption{Comparison of Queyranne's algorithm with REMCMC in ECoG data (SI) }\label{Tab:QvsREMCMC_SI_NeuroTycho}
  \begin{tabular}{c c c  c c c} \toprule
     \multicolumn{3}{c}{\textbf{Winning percentage}} & \multicolumn{3}{c}{\textbf{Number of evaluations of $\Phi$}} \\ \cmidrule(lr){1-3} \cmidrule(lr){4-6}
                  &       &        &             & \multicolumn{2}{c}{REMCMC (mean$\pm$std)} \\ \cmidrule(lr){5-6}
      Queyranne's &  Even & REMCMC & Queyranne's &           Converged &           Solution found \\ \midrule
              0\% & 100\% &    0\% &      87,423 & 607,797$\pm$410,588 & 15,859$\pm$10,497 \\ \bottomrule
  \end{tabular}
  \end{center}
\end{table}

\section{Discussions}\label{Sec:Discussions}
In this study, we proposed an efficient algorithm for searching for the Minimum Information Partition (MIP) in Integrated Information Theory (IIT). The computational time of an exhaustive search for the MIP grows exponentially with the arithmetic growth of system size, which has been an obstacle to applying IIT to experimental data. We showed here that by using a submodular optimization algorithm called Queyranne's algorithm, the computational time was reduced to $O(N^{3.066})$ and $O(N^{4.776})$ for stochastic interaction $\Phi_\mathrm{SI}$ and geometric integrated information $\Phi_\mathrm{G}$, respectively. These two measures of integrated information are non-submodular, and thus it is not theoretically guaranteed that Queyranne's algorithm will find the MIP. We empirically evaluated the accuracy of the algorithm by comparing it with an exhaustive search in simulated data and in ECoG data recorded from monkeys. We found that Queyranne's algorithm worked perfectly for $\Phi_\mathrm{SI}$ and almost perfectly for $\Phi_\mathrm{G}$. We also tested the performance of Queyranne's algorithm in larger systems ($N= 20$ and 50 for $\Phi_\mathrm{SI}$ and $\Phi_\mathrm{G}$, respectively) where the exhaustive search is intractable by comparing it with the Replica Exchange Markov Chain Monte Carlo method (REMCMC). We found that the partitions found by these two algorithms perfectly matched, which suggests that both algorithms most likely found the MIPs. In terms of the computational time, the number of evaluations of $\Phi$ taken by Queyranne's algorithm was much smaller than that taken by REMCMC before the convergence. Our results indicate that Queyranne's algorithm can be utilized to effectively estimate MIP even for non-submodular measures of integrated information. 
\Com{Although the MIP is a concept originally proposed in IIT for understanding consciousness, it can be utilized to general network analysis  irrespective of consciousness. Thus, the method for searching MIP proposed in this study will be beneficial not only for consciousness studies but for other research fields.}

Here, we discuss the pros and cons of Queyranne's algorithm in comparison with REMCMC. Since the partitions found by both algorithms perfectly matched in our experiments, they were equally good in terms of accuracy. With regards to computational time, Queyranne's algorithm ended much faster than the convergence of REMCMC. Thus, Queyranne's algorithm would be a better choice in rather large systems ($N\sim 20$ and 50 for $\Phi_\mathrm{SI}$ and $\Phi_\mathrm{G}$, respectively). Note that if we retrospectively examine the sampling sequence in REMCMC, we find that REMCMC found the partitions much earlier than its convergence and that the estimated MIPs did not change in the later parts of sampling process. Thus, if we could introduce a heuristic criterion to determine when to stop the sampling based on the time course of the estimated MIPs, REMCMC could be stopped earlier than its convergence. However, setting such a heuristic criterion is a non-trivial problem. Queyranne's algorithm ends within a fixed number of function calls regardless of the properties of data. If the system size is much larger ($N \gtrsim 100$), Queyranne's algorithm will be computationally very demanding because of $O(N^3)$ time complexity and may not practically work. In that case, REMCMC would work better if the above-mentioned heuristics are introduced to stop the algorithm earlier than the convergence.

\Com{As an alternative interesting approach for approximately finding the MIP, a graph-based algorithm was proposed by Toker \& Sommer \cite{Toker2017}. In their method, to reduce the search space, candidate partitions are selected by a spectral clustering method based on correlation. Then $\Phi_\mathrm{G}$ is calculated for those candidate partitions, and the best partition is selected. A difference between our method and theirs is whether the search method is fully based on the values of integrated information or not. Our method uses no other quantities than $\Phi$ for searching the MIP, while their method uses a graph theoretic measure, which may significantly differ from $\Phi$ in some cases. It would be an interesting future work to compare our method and the graph-theoretic methods or combine these methods to develop better search algorithms.}

In this study,
\Com{we considered the three different measures of integrated information, $\Phi_\mathrm{MI}$, $\Phi_\mathrm{SI}$, and $\Phi_\mathrm{G}$. Of these, $\Phi_\mathrm{MI}$ is submodular but the other two measures, $\Phi_\mathrm{SI}$ and $\Phi_\mathrm{G}$, are not.}
As we described in Section \ref{seq:sub_phi}, there is a clear order relation among them (Eq. (\ref{Eq:order})). $\Phi_\mathrm{SI}$ is closer to a submodular function $\Phi_\mathrm{MI}$ than $\Phi_\mathrm{G}$ is. This relation implies that Queyranne's algorithm would work better for $\Phi_\mathrm{SI}$ than for $\Phi_\mathrm{G}$. We found that it was actually the case in our experiments because there were a few error trials for $\Phi_\mathrm{G}$ whereas there were no miss trials for $\Phi_\mathrm{SI}$. For the practical use of these measures, we note that there are two major differences among the three measures. One is what they quantify. As shown in Fig. \ref{fig:phis}, $\Phi_\mathrm{G}$ measures only causal interactions between units across different time points. In contrast, $\Phi_\mathrm{SI}$ and $\Phi_\mathrm{MI}$ also measure equal time interactions as well as causal interactions. $\Phi_\mathrm{G}$ best follows the original concept of IIT in the sense that it measures only the ``causal'' interactions. One needs to acknowledge the theoretical difference whenever applying one of these measures in order to correctly interpret the obtained results. The other difference is in computational costs. The computational costs of $\Phi_\mathrm{MI}$ and $\Phi_\mathrm{SI}$ are almost the same while that of $\Phi_\mathrm{G}$ is much larger, because it requires multi-dimensional optimization. Thus, $\Phi_\mathrm{G}$ may not be practical for the analysis of large systems. In that case, $\Phi_\mathrm{MI}$ or $\Phi_\mathrm{SI}$ may be used instead with care taken of the theoretical difference. 

\Com{Although in this study we focused on bi-partitions, Queyranne’s algorithm can be extended to higher-order partitions \cite{Hidaka2017}. However, the algorithm becomes computationally demanding for higher-order partitions, because the computational complexity of the algorithm for $K$-partitions is $O(N^{3(K-1)})$. This is the main reason why we focused on bi-partitions. Another reason is that there has not been an established way to fairly compare partitions with different $K$. In IIT 2.0, it was proposed that the integrated information should be normalized by the minimum of the entropy of partitioned subsystems \cite{Tononi2008}, while in IIT 3.0, it was not normalized \cite{Oizumi2014}. Note that when integrated information is not normalized, the MIP is always found in bi-partitions because integrated information becomes larger when a system is partitioned into more subsystems.}

\Com{Whether the integrated information should be normalized and how the integrated information should be normalized are still open questions. In our study, the normalization used in IIT 2.0 is not appropriate, because the entropy can be negative for continuous random variables. Additionally, regardless of whether random variables are continuous or discrete, normalization significantly affects the submodularity of the measures of integrated information. For example, if we use normalization proposed in IIT 2.0, even the submodular measure of integrated information, $\Phi_\mathrm{MI}$, no longer satisfies submodularity. Thus, Queyranne’s algorithm may not work well if $\Phi$ is normalized.}


Although we resolved one of the major computational difficulties in IIT, an additional issue still remains. Searching for the MIP is an intermediate step in identifying the informational core, called the ``complex''. The complex is the subnetwork in which integrated information is maximized, and is hypothesized to be the locus of consciousness in IIT. Identifying the complex is also represented as a discrete optimization problem which requires exponentially large computational costs. Queyranne's algorithm cannot be applied to the search for the complex because we cannot formulate it as a submodular optimization. We expect that REMCMC would be efficient in searching for the complex and will investigate its performance in a future study. 

An important limitation of this study is that we only showed the nearly perfect performance of Queyranne's algorithm in limited simulated data and real neural data. In general, we cannot tell whether Queyranne's algorithm works well for other data beforehand. For real data analysis, we recommend that the procedure below should be applied. First, as we did in Section \ref{Subsec:Accuracy}, accuracy should be checked by comparing it with the exhaustive search in small randomly selected subsets. Next, if it works well, the performance should be checked by comparing it with REMCMC in relatively large subsets, as we did in Section \ref{Subsec:QvsREMCMC}. If Queyranne's algorithm works better than or equally as well as REMCMC, it is reasonable to use Queyranne's algorithm for the analysis. By applying this procedure, we expect that Queyranne's algorithm could be utilized to efficiently find the MIP in a wide range of time series data. 

\vspace{6pt} 

\acknowledgments{We thank Dr. Shohei Hidaka, Japan Advanced Institute of Science and Technology, for providing us Queyranne’s algorithm codes. This work was partially supported by JST CREST Grant Number JPMJCR15E2, Japan.}

\authorcontributions{
J.K. and M.O. conceived and designed the experiments; J.K. performed the experiments; J.K. and M.O. analyzed the data; and J.K., R.K. and M.O. wrote the paper.
}

\conflictofinterests{The authors declare no conflict of interest. The founding sponsors had no role in the design of the study; in the collection, analyses, or interpretation of data; in the writing of the manuscript, or in the decision to publish the results.}

\abbreviations{The following abbreviations are used in this manuscript:\\

\noindent IIT: integrated information theory\\
MIP: minimum information partition\\
MCMC: Markov chain Monte Carlo\\
REMCMC: replica exchange Markov chain Monte Carlo\\
ECoG: electrocorticogram\\
AR: autoregressive\\
CR: correct rate\\
RA: rank\\
ER: error ratio\\
CORR: correlation\\
MCS: Monte Carlo step
}


\appendix
\section{Analytical formula of $\Phi$ for Gaussian variables}\label{app:Phi_Gauss}
We describe the analytical formula of \Com{three} measures of integrated information, \Com{multi information ($\Phi_\mathrm{MI}$),} stochastic interaction ($\Phi_\mathrm{SI}$) and geometric integrated information ($\Phi_\mathrm{G}$), when the probability distribution is Gaussian. For more details about the theoretical background, see \cite{Oizumi2016,Barrett2010,Ay2001,Ay2015}.

First, let us introduce the notation. We consider a stochastic dynamical system consisting of $N$ elements. We represent the past and present states of the system as $X=(X_1, \ldots , X_N)$ and $X'=(X'_1, \ldots , X'_N)$, respectively, and define a joint vector
\begin{equation}
\tilde{X}=(X, X').
\end{equation}
We assume that the joint probability distribution $p\left(X,X'\right)$ is Gaussian: 
\begin{equation}
  p\left( x,x' \right) = \exp \left\{ -\frac{1}{2} \left(\tilde{x}^T \Sigma(\tilde{X})\tilde{x} -\psi \right) \right\}, \label{eq:GaussDist}
\end{equation}
where $\psi$ is the normalizing factor and $\Sigma(\tilde{X})$ is the covariance matrix of $\tilde{X}$. 
\Com{Note that we can assume the mean of the Gaussian distribution is zero without loss of generality because the mean value does not affect the values of integrated information.}
This covariance matrix $\Sigma(\tilde{X})$ is given by 
\begin{equation}
  \Sigma(\tilde{X}) = \begin{pmatrix}
\Sigma(X)       & \Sigma(X, X')\\
\Sigma(X, X')^T & \Sigma(X') 
\end{pmatrix},
\end{equation}
where $\Sigma(X)$ and $\Sigma(X')$ are the equal time covariance at past and present, respectively, and $\Sigma(X, X')$ is the cross covariance between $X$ and $X'$. 
Below we will show the analytical expression of \Com{$\Phi_\mathrm{MI}$,} $\Phi_\mathrm{SI}$ and $\Phi_\mathrm{G}$.  

\Com{
\subsection{Multi information}
Let us consider the following partitioned probability distribution $q$, 
\begin{equation}
  q\left(X,X'\right)=\prod_i q\left(M_i,M'_i\right),
\end{equation}
where $M_i$ and $M_i'$ are the past and present states of $i$-th subsystem. Then multi information is defined as 
\begin{equation}
\Phi_\mathrm{MI} = \sum_i H(M_i,M'_i) - H(X,X'). \label{eq:Phi_MI}
\end{equation}
When the distribution is Gaussian, Eq. (\ref{eq:Phi_MI}) is transformed to 
\begin{equation}
  \Phi_\mathrm{MI} = \sum_i \log|\Sigma(\tilde{M}_i)| - \log|\Sigma(\tilde{X})|, 
\end{equation}
where $\tilde{M}_i = (M_i,M'_i)$ and $\Sigma(\tilde{M}_i)$ is the covariance of $\tilde{M}_i$. 
}

\subsection{Stochastic interaction}
We consider the following partitioned probability distribution $q$, 
\begin{equation}
  q\left(X'|X\right)=\prod_i q\left(M'_i|M_i\right).
\end{equation}
Then stochastic interaction \cite{Oizumi2016,Barrett2010,Ay2001,Ay2015} is defined as 
\begin{equation}
\Phi_\mathrm{SI} = \sum_i H(M'_i|M_i) - H(X'|X). \label{eq:Phi_SI}
\end{equation}
When the distribution is Gaussian, Eq. (\ref{eq:Phi_SI}) is transformed to 
\begin{equation}
  \Phi_\mathrm{SI} = \sum_i \log |\Sigma(M_i'|M_i)| - \log|\Sigma(X'|X)|, 
\end{equation}
where $\Sigma(M_i'|M_i)$ and $\Sigma(X'|X)$ are covariance matrices of conditional distributions. These matrices are represented as
\begin{equation}
  \begin{split}
    \Sigma(M_i'|M_i) &= \Sigma(M_i')-\Sigma(M_i, M_i')^T \Sigma(M_i)^{-1} \Sigma(M_i, M_i'), \\
    \Sigma(X'|X) &= \Sigma(X')-\Sigma(X, X')^T \Sigma(X)^{-1} \Sigma(X, X'),
  \end{split}
\end{equation}
where $\Sigma(M_i)$ and $\Sigma(M_i')$ are the equal time covariance of subsystem $i$ at past and present, respectively, and $\Sigma(M_i, M_i')$ is the cross covariance between $M_i$ and $M_i'$. 

\subsection{Geometric integrated information}
To calculate the geometric integrated information \cite{Oizumi2016}, we first transform Eq. (\ref{eq:GaussDist}). 
Equation \ref{eq:GaussDist} is equivalently represented as an autoregressive model: 
\begin{equation}
 X' = A X + E, 
\end{equation}
where $A$ is the connectivity matrix and $E$ is Gaussian random variables, which are uncorrelated over time. 
By using this autoregressive model, the joint distribution $p\left( X,X' \right)$ is expressed as 
\begin{equation}
  p\left( x,x' \right) = \exp \left\{ -\frac{1}{2} \left(
  x^T \Sigma(X) x + (x'-Ax)^T\Sigma(E)^{-1}(x'-Ax)  -\psi \right) \right\}, 
\end{equation}
and the covariance matrices as 
\begin{equation}
 \begin{split}
   \Sigma(X,X') &= \Sigma(X)A^T, \\
   \Sigma(X') &= \Sigma(E) + A\Sigma(X)A^T, 
 \end{split}
\end{equation}
where $\Sigma(E)$ is the covariance of $E$. 
Similarly, the joint probability distribution in a partitioned model is given by
\begin{equation}
  \begin{split}
  q\left( x,x' \right) &= \exp \left\{ -\frac{1}{2} \left(\tilde{x}^T \Sigma(\tilde{X})_\mathrm{p}\tilde{x} -\psi \right) \right\}\\
  &=\exp \left\{ -\frac{1}{2} \left( x^T \Sigma(X)_\mathrm{p} x + (x'-A_\mathrm{p}x)^T\Sigma(E)_\mathrm{p}^{-1}(x'-A_\mathrm{p}x)  -\psi \right) \right\},
  \end{split}
\end{equation}
where $\Sigma(X)_\mathrm{p}$ and $\Sigma(E)_\mathrm{p}$ are the covariance matrices of $X$ and $E$ in the partitioned model, respectively, and $A_\mathrm{p}$ is the connectivity matrix in the partitioned model. 

The geometric integrated information is defined as 
\begin{align}
\Phi_\mathrm{G} &= \min_{q} D_\mathrm{KL}\left(p\left(X,X'\right)||q\left(X,X'\right)\right), \label{eq:Phi_G} \\ 
D_\mathrm{KL}\left(p\left(X,X'\right)||q\left(X,X'\right)\right) &= \frac{1}{2} \left( \log\frac{|\Sigma(\tilde{X})_\mathrm{p}|}{|\Sigma(\tilde{X})|} + \mathrm{Tr}(\Sigma(\tilde{X}) \Sigma(\tilde{X}_\mathrm{p})^{-1})-2N \right), \label{eq:D_KL_Phi_G}
\end{align}
such that
\begin{equation}
q\left(M'_i|X\right)=q\left(M'_i|M_i\right), \forall i. \label{eq:const_Phi_G}
\end{equation}
This constraint (Eq. (\ref{eq:const_Phi_G})) corresponds to setting the between-subsystem blocks of $A_\mathrm{p}$ to 0: 
\begin{equation}
  (A_\mathrm{p})_{ij} = 0 \; (i\neq j).
\end{equation}
By transforming stationary point conditions, $\partial D_\mathrm{KL}/\partial \Sigma(\tilde{X})_\mathrm{p}^{-1}=0$, $\partial D_\mathrm{KL}/\partial(A_\mathrm{p})_{ii}=0$, and $\partial D_\mathrm{KL} / \partial\Sigma(E)_\mathrm{p}^{-1}=0$, we get 
\begin{gather}
  \Sigma(X)_\mathrm{p} = \Sigma(X), \label{eq:cont_Phi_G_SigmaX}\\
  (\Sigma(X)(A-A_\mathrm{p})\Sigma(E)_\mathrm{p}^{-1})_{ii}=0, \label{eq:cont_Phi_G_A}\\
  \Sigma(E)_\mathrm{p} = \Sigma(E) + (A-A_\mathrm{p})\Sigma({X})(A-A_\mathrm{p})^T. \label{eq:cont_Phi_G_SigmaE}
\end{gather}
By substituting Eqs. (\ref{eq:cont_Phi_G_SigmaX}) and  (\ref{eq:cont_Phi_G_SigmaE}) into Eq. (\ref{eq:Phi_G}), $\Phi_\mathrm{G}$ is simplified as 
\begin{equation}
  \Phi_\mathrm{G} = \frac{1}{2}\log \frac{|\Sigma(E)_\mathrm{p}|}{|\Sigma(E)|}. \label{eq:Phi_G_final} 
\end{equation}
\Com{To obtain the value of Eq. (\ref{eq:Phi_G_final}), we need to find the value of $\Sigma(E)_\mathrm{p}$, which requires solving Eqs. (\ref{eq:cont_Phi_G_SigmaX}), (\ref{eq:cont_Phi_G_A}) and (\ref{eq:cont_Phi_G_SigmaE}). Thus, the calculation of $\Phi_\mathrm{G}$ requires solving multi-dimensional equations.}
The MALAB codes for this computation of $\Phi_\mathrm{G}$ are available at \cite{Kitazono2017}.

\section{Details of Replica Exchange Markov Chain Monte Carlo Method}\label{app:REMCMC}
The Replica Exchange Markov Chain Monte Carlo (REMCMC) method was originally proposed to investigate physical systems \cite{Swendsen1986replica,Geyer1991markov,Hukushima1996exchange}, and was then rapidly utilized in other applications, including combinatorial optimization problems \cite{Pinn1998number,Hukushima2002extended,Barthel2004,Wang2009,Nagata2015exhaustive}. For a more detailed history of REMCMC, see, for example, \cite{Earl2005parallel}.

We first briefly explain how the MIP search problem is dealt with by the Metropolis method. Then, as an improvement of Metropolis method, we introduce REMCMC to more effectively search for the global minimum while avoiding being trapped around at a local minimum. Next, we describe the convergence criterion of MCMC sampling. Finally we present the parameter settings in our experiments. 

\subsection{Metropolis method}

We consider the way to sample subsets from the probability distribution Eq. (\ref{eq:Boltzmann}). An initial subset $S^{(0)}$ is randomly selected, and then a sample sequence is drawn as follows. 
\begin{description}
\item[Propose a candidate of the next sample]
An element $e$ is randomly selected and if it is in the current subset $S^{(t)}$, the candidate $S_\mathrm{c}$ is $S^{(t)}\setminus\{e\}$. If not, the candidate is $S^{(t)}\cup \{e\}$. 

\item[Determine whether to accept the candidate or not]
The candidate $S_\mathrm{c}$ is accepted ($S^{(t+1)}=S_\mathrm{c}$) or not accepted ($S^{(t+1)}=S^{(t)}$) according to the following probability $a(S^{(t)}\rightarrow S_\mathrm{c})$: 
\begin{equation}
  \begin{split}
    a(S^{(t)}\rightarrow S_\mathrm{c}) &= \min(1,r), \\
    r &= \frac{p(S_\mathrm{c};\beta)}{p(S^{(t)};\beta)}= \exp\left[ \beta \left\{\Phi(S^{(t)})-\Phi(S_\mathrm{c}) \right\} \right]. \label{eq:Metropolis}
  \end{split}
\end{equation}
This probability means that if the integrated information decreases by stepping from $S^{(t)}$ to $S_\mathrm{c}$, the candidate $S_\mathrm{c}$ is always accepted, and otherwise it is accepted with the probability $r$. 

\end{description}
By iterating these two steps with sufficient time, the sample distribution converges to the probability distribution given in Eq. (\ref{eq:Boltzmann}). 
$N$ steps of the sampling is referred to as one Monte Carlo step (MCS), where $N$ is the number of elements. In one MCS, each element is attempted to be added or removed once on average. 

Depending on the value of $\beta$, the behavior of the sample sequence changes. If $\beta$ is small, the probability distribution given by Eq. (\ref{eq:Boltzmann}) is close to a uniform distribution and subsets are sampled nearly independently of the value of $\Phi(S)$. If $\beta$ is large, the candidate is more likely to be accepted when the integrated information decreases. The sample sequence easily falls to a local minimum and cannot explore many subsets. Thus, smaller and larger $\beta$ have an advantage and a disadvantage: Smaller $\beta$ is better for exploring around many subsets while larger $\beta$ is better for finding a (local) minimum. In the Metropolis method, we need to set $\beta$ to an appropriate value taking account of this trade-off, but it is generally difficult.

\subsection{Replica Exchange Markov chain Monte Carlo}
To overcome the difficulty in setting inverse temperature $\beta$, REMCMC samples from distributions at multiple values of $\beta$ in parallel and the sampled sequences are exchanged between nearby values of $\beta$. By this exchange, the sampled sequences at high inverse temperatures can escape from local minima and can explore many subsets. 

We consider $M$-probabilities at different inverse temperatures $\beta_1>\beta_2>\cdots>\beta_M$ and introduce the following joint probability:
\begin{equation}
  p(S_1,\ldots,S_M;\beta_1\ldots\beta_M) = \prod_{m=1}^{M}p(S_m;\beta_m). \label{eq:jointp}
\end{equation}
Then, the simulation process of the REMCMC consists of the following two steps. 
\begin{description}
\item[Sampling from each distribution]
Samples are drawn from each distribution $p(S_m;\beta_m)$ separately by using the Metropolis method as described in the previous subsection. 
\item[Exchange between neighboring inverse temperatures]
After a given number of samples are drawn, subsets at neighboring inverse temperatures are swapped, according to the following probability $p(S_m\leftrightarrow S_{m+1})$:
\begin{equation}
  \begin{split}
    p(S_m\leftrightarrow S_{m+1}) &= \min(1,r'), \\ \label{eq:ExchangeProb}
    r' &= \frac{p(S_{m+1};\beta_m)p(S_m;\beta_{m+1})}{p(S_m;\beta_m)p(S_{m+1};\beta_{m+1})} \\
       &= \exp\left[ (\beta_{m+1}-\beta_m) \left\{\Phi(S_{m+1})-\Phi(S_m) \right\} \right].
  \end{split}
  \end{equation}
This probability indicates that if the integrated information at a higher inverse temperature is larger than that at a lower inverse temperature, subsets are always swapped; and otherwise, they are swapped with the probability $r'$.

\end{description}
By iterating these two steps for sufficient time, the sample distribution converges to the joint distribution Eq.~(\ref{eq:jointp}).

To maximize the efficiency of the REMCMC, it is important to appropriately set the multiple inverse temperatures. If the neighboring temperatures are far apart, the acceptance ratio of exchange (Eq.~(\ref{eq:ExchangeProb})) becomes too small. The REMCMC is then reduced to just separately simulating distributions at different temperatures without any exchange. In a previous study \cite{Rathore2005optimal}, it was recommended to keep the average ratio higher than 0.2 for every temperature pair. At the same time, the highest/lowest inverse temperatures should be high/low enough so that sample sequence at the highest inverse temperature can reach the tips of (local) minima and that at the lowest one can search around many subsets.
To satisfy these constraints, a sufficient number $M$ of inverse temperatures are accommodated and the inverse temperatures are optimized to equalize the average of the acceptance ratio of exchanges at all temperature pairs \cite{Sugita1999141,Kofke2002,Kofke2004Erratum,Rathore2005optimal,Lee2011comparison}. Details of temperature setting are described below. 

\paragraph{Initial setting}
Inverse temperatures $\beta_m (m=1,\ldots,M)$ are initially set as follows. First, a subset is randomly selected for each $m$. Then, a randomly chosen element is added to or eliminated from each subset, and the absolute value of the change $\Delta\Phi_m$ in the amount of integrated information is taken. By using these absolute values, the highest and lowest inverse temperatures are determined by a bisection method so that the respective averages of the acceptance ratio $\exp(-\beta\Delta\Phi_1)$ and $\exp(-\beta\Delta\Phi_M)$ match the predefined values. The intermediate inverse temperatures are set to be a geometric progression: $\beta_m = \beta_1\left(\frac{\beta_M}{\beta_1}\right)^{\frac{m-1}{M-1}}$. 

\paragraph{Updating}
The difference in the amount of integrated information between the candidate subset $\Phi(S_\mathrm{c})$ and the current subset $\Phi(S^{(t)})$ is stored when the difference is positive ($\Phi(S_\mathrm{c})-\Phi(S^{(t)})\geq0$). Then, by using the stored values at all the inverse temperatures, the highest and lowest inverse temperatures are determined by a bisection method so that the average of the acceptance ratio $\exp\left[ \beta \left\{\Phi(S^{(t)})-\Phi(S_\mathrm{c}) \right\} \right]$ matches the predefined value, as in the initial setting. 
The intermediate inverse temperatures are set to approximately equalize the expected values of acceptance ratio of the exchange at all temperature pairs \cite{Sugita1999141,Kofke2002,Kofke2004Erratum,Rathore2005optimal,Lee2011comparison}. 
The expected value is represented as a sum of two probabilities:
\begin{equation}
  \begin{split}
  \mathbb{E}\left[ p(S_m\leftrightarrow S_{m+1}) \right] = & \int_{-\infty}^\infty \int_{-\infty}^\infty \bigg\{ p(\Phi_m\geq\Phi_{m+1})\\
  & \left. + p(\Phi_m<\Phi_{m+1})e^{(\beta_m-\beta_{m+1})(\Phi_m-\Phi_{m+1})} \right\}\mathrm{d}\Phi_m\mathrm{d}\Phi_{m+1}.
  \end{split}
\end{equation}
In \cite{Lee2011comparison}, 
this expected value is approximated as 
\begin{equation}
  \begin{split}
  \mathbb{E}\left[ p(S_m\leftrightarrow S_{m+1}) \right] \approx &\frac{1}{2} \mathrm{erfc}\left( \frac{\mu(T_{m+1})-\mu(T_{m})}{\sqrt{ 2\left\{\sigma^2(T_{m+1})+\sigma^2(T_m) \right\} }} \right)\\
    &+ \left\{ 1 - \frac{1}{2}\mathrm{erfc}\left( \frac{\mu(T_{m+1})-\mu(T_{m})}{\sqrt{2\left\{\sigma^2(T_{m+1})+\sigma^2(T_m) \right\} }} \right)  \right\} e^{(\beta_m-\beta_{m+1})(\mu(T_m)-\mu(T_{m+1}))},
  \end{split}
\end{equation}
where $\mu(T)$ and $\sigma^2(T)$ are the mean and variance of $\Phi$, represented as functions of temperature $T$. 
In \cite{Lee2011comparison}, these functions are given by interpolating the sample mean and variance. 
In this study, these functions are estimated using regression, because the sample mean and variance are highly variable. The mean and variance at each temperature are computed at every update, and these means and variances are regressed on temperature using a continuous piecewise linear function, the $T$-axis of anchor points of which are current temperatures. The anchor points are interpolated using piecewise cubic Hermite interpolating polynomials.  
Then, to roughly equalize the expected values of the acceptance ratio of the exchange at all temperature pairs, we minimize the following cost function by varying temperatures \cite{Lee2011comparison}: 
\begin{equation}
  \mathrm{Cost} = \sum_{m=1}^{M-1} \mathbb{E}\left[ p(S_m\leftrightarrow S_{m+1}) \right]^{-4}.
\end{equation}
The minimization is performed by a line-search method.

\subsection{Convergence criterion}\label{app:convergence}
One of the most commonly used MCMC convergence criteria is potential scale reduction factor (PSRF), which was proposed by Gelman \& Rubin (1992) \cite{gelman1992inference}, and modified by Brooks \& Gelman (1998) \cite{brooks1998general}. 
In this criterion, multiple MCMC sequences are run. If all of them converge, statistics of the sequences must be about the same. This is assessed by comparing between-sequence variance and within-sequence variance of a random variable and calculating the PSRF, $\hat{R}_c$. Large $\hat{R}_c$ suggests that some of the sequences do not converge yet. If $\hat{R}_c$ is close to 1, we can diagnose them as converged. 
In this study, we cut the sequence at each inverse temperature into the former and the latter halves, and applied the criterion to these two half sequences. If $\hat{R}_c$ of all the temperatures were below a predefined threshold, we regarded the sequences as converged. 

\subsection{Parameter settings}
The number of inverse temperatures $M$ was fixed at 6 throughout out the experiments. 
The highest/lowest inverse temperatures were set so that the averages of acceptance ratio become 0.01 and 0.5, respectively. The exchange process was done every 5 MCSs. The update of inverse temperatures was performed every 5 MCSs for the 200 initial MCSs. The threshold of $\hat{R}_c$ was set to 1.01. When computing $\hat{R}_c$, we discarded the first 200 MCSs as a burn-in period and started to computing it after 300 MCSs. 

\Com{
\section{Values of $\Phi$}\label{app:boxplots}
\begin{figure}[h]
 \begin{center}
  \includegraphics[width=1\hsize]{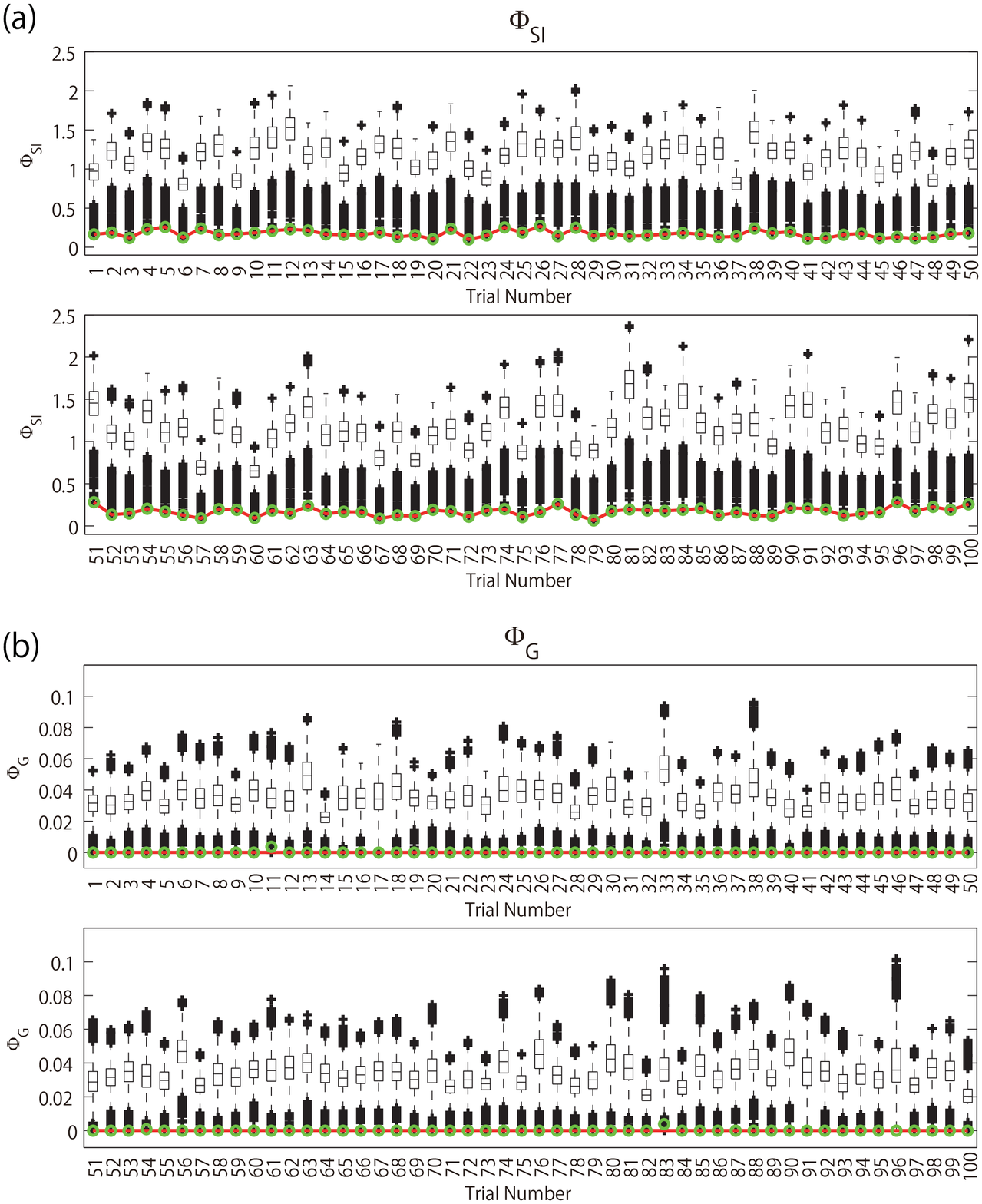} 
  \caption{The values of $\Phi$ for the block-structured models at $\sigma=0.01$. The box plots represent the distribution of $\Phi$ at all the partitions. The red solid line indicates $\Phi$ at the MIP. The green circles indicate $\Phi$ at the partitions found by Queyranne's algorithm. (a) $\Phi_\mathrm{SI}$, (b) $\Phi_\mathrm{G}$.} \label{fig:boxplots_SI_G_block}
 \end{center}
\end{figure}
We show some examples of the distributions of the values of $\Phi$ in the experiments in Subsection \ref{Subsec:Accuracy}. Figures \ref{fig:boxplots_SI_G_block} (a) and (b)  are the box plots of $\Phi_\mathrm{SI}$ and $\Phi_\mathrm{G}$ for the block-structured models at $\sigma=0.01$, respectively. We can see that in Fig. \ref{fig:boxplots_SI_G_block} (a), $\Phi_\mathrm{SI}$ computed at the partition found by Queyranne's algorithm perfectly matched with that at the MIPs. In Fig. \ref{fig:boxplots_SI_G_block} (b), $\Phi_\mathrm{G}$ computed at the partition found by Queyeranne's algorithm did not match that at the MIPs in 3 trials (the trial numbers 11, 54 and 83) but the deviations were very small.}

\bibliographystyle{mdpi}
\bibliography{MIP}
\end{document}